\documentstyle[prl,aps,epsf,rotate]{revtex}
\draft

\begin{document}
%%% Prints in two-column form
\twocolumn[\hsize\textwidth\columnwidth\hsize
     \csname @twocolumnfalse\endcsname

\title{A dynamic model of atoms:\\
structure, internal interactions and photon emissions of hydrogen}

\author{W. A. Hofer}
\address{
         Dept. of Physics and Astronomy, University College London \\
         Gower Street, London WC1E 6BT, E-mail: w.hofer@ucl.ac.uk}

\maketitle

%---------------------------------------------------------------------------%
%       A B S T R A C T    A N D   P A C S                                  %
%---------------------------------------------------------------------------%

\begin{abstract}
    The standard solution of the Schr\"odinger equation for the
    hydrogen atom is analyzed. Comparing with the recently established
    internal properties of electrons it is found,
    that these solutions cannot be seen as physically valid states of the 
    electron wave.    
    The paper therefore proposes a new model of hydrogen based on
    internal properties of electrons. The ground state of the
    hydrogen system (T=0) is an inertial aggregation within the 
    atomic shell, the calculation yields an atomic radius of 
    $0.330$ nm. Electron proton interaction within the atom
    are treated with a causal and deterministic model, the
    resonance frequency of the hydrogen system $ \nu_{0} =
    6.57 \times 10^{15} Hz $ is referred to dynamic charge of
    its nucleus, resonance levels are a result
    of boundary conditions for radial electron waves and photon
    interactions due to nuclear oscillations.
    Spectral emissions of excited atoms can be referred to a decay
    of the state of motion of the coupled electron--proton system.
    The framework developed is essentially deterministic, microphysical
    processes analyzed are referred to material characteristics of 
    particles involved. Statistical effects are referred to interactions
    with the atomic environment, the results derived are
    compatible with the second and third principle of thermodynamics.
\end{abstract} 


%%% end of two-column print
\vskip2pc]

%---------------------------------------------------------------------------%
%      SECTION 1:  I N T R O D U C T I O N                                  %
%---------------------------------------------------------------------------%

\section{Introduction}

In recent publications \cite{HOF97A,HOF98A,HOF00A,HOF00B} we
demonstrated, that fundamental relations of quantum theory have
to be understood as a consequence of intrinsic particle
structures. The discovery of intrinsic potentials, inherent
to wave properties of moving particles described by the 
modified de Broglie relations, allowed to deduce the fundamental
relations of quantum theory as well as electrodynamics, and it
was shown, that these intrinsic properties do possess
physical significance.

The Schr\"odinger equation \cite{SCH26} was identified as a 
modification of the wave equation including external potentials,
and it was demonstrated that in its traditional formulation it
does not account for intrinsic potentials, which makes its
fundamental results arbitrary. Heisenberg's uncertainty relations
were deduced by estimating the effect of this arbitrariness.

We could equally show, that the interaction of electrons and 
photons accounts for the observed quantization, although
quantization must be understood as a result of transfer processes at
every single point of a given region of interaction. The analysis
of electrostatic interactions finally allowed the conclusion
that electrostatic fields are determined by photon interactions.

In this paper we analyze the traditional model of
hydrogen atoms derived from solutions of Schr\"odinger's equation
including a central electrostatic potential, and develop the 
consequences for material waves. Based on the results of this 
analysis the experimental constraints are determined, which
must be observed in the construction of an alternative model
compatible with the intrinsic wave features of electrons. 
Finally, by formalizing the process of photon emission, we derive
a deterministic theory of electron states within the hydrogen
atom, which includes the electromagnetic fields of emission.

%---------------------------------------------------------------------------%
%      SECTION 2:  C U R R E N T   S T A N D A R D                    %
%---------------------------------------------------------------------------%

\section{Current standard}\label{sec_qt}

The atomic structure, as well as the, then puzzling, experimental
results in atomic physics were the main motivation for  
constructing the theoretical framework of quantum theory.
It was, basically, the only way out of the profound dilemma
of physics after the first attempts by Bohr and Sommerfeld --
today known as the ''old quantum theory'' \cite{BOHR13,SOM16} -- 
failed, to explain measurements by a model consistent with 
classical mechanics enhanced by Einstein's concept of photons
\cite{EIN05A}.

\subsection{Solutions of Schr\"odinger's equation}

Limiting calculations of hydrogen ground--states to the Hamiltonian
of electron motion in the field of a central charge +e, not
accounting for spin--properties and electron--proton interactions
as well as neglecting applied electrostatic or magnetic fields,
the Schr\"odinger equation reads \cite{SCH26}:

\begin{eqnarray}\label{qt001}
        \left( - \frac{\hbar^2}{2 m} \triangle - 
        \frac{e^2}{r} \right) \psi (\vec r) = 
        E \psi (\vec r)
\end{eqnarray}

The standard procedure separates the radial and lateral
components of $\psi(\vec r)$ (see, for example \cite{MES69}).

\begin{eqnarray} \label{qt002}
\psi(\vec r) = \frac{y_{l}(r)}{r}\,Y_{l}^{m}(\vartheta, \varphi)
\end{eqnarray}

Separating the Laplace operator in spherical coordinates:

\begin{eqnarray}\label{qt003}
        \triangle &=& \frac{1}{r^2} \frac{\partial}{\partial r} 
        r^2 \frac{\partial}{\partial r}
        + \frac{1}{r^2 \sin \vartheta} 
        \left(\frac{\partial}{\partial \vartheta} \sin \vartheta
        \frac{\partial}{\partial \vartheta} +
        \frac{1}{\sin \vartheta} 
        \frac{\partial^2}{\partial \varphi^2} \right)\nonumber
\end{eqnarray}

using the eigenvalue equation for the spherical harmonics 
$Y_{l}^{m}(\vartheta,\varphi)$:

\begin{eqnarray}\label{qt004}
        \frac{1}{\sin \vartheta} 
        \left(\frac{\partial}{\partial \vartheta} \sin \vartheta 
        \frac{\partial}{\partial \vartheta} +
        \frac{1}{\sin \vartheta} 
        \frac{\partial^2}{\partial \varphi^2} \right)
        Y_{l}^{m} = l(l + 1) Y_{l}^{m}
\end{eqnarray}

the Schr\"odinger equation leads to the radial equation:

\begin{eqnarray}\label{qt005}
        y_{l}'' + \left(\frac{2 m E}{\hbar^2} +
        \frac{2 m}{\hbar^2} \frac{e^2}{r} - 
        \frac{l(l + 1)}{r^2} \right) y_{l} = 0
\end{eqnarray}

A substitution of variables 
$x = 2r \sqrt{- \frac{2 m E}{\hbar^2}}$ yields for 
the regular solution of $y(x)$ the expression:

\begin{eqnarray}\label{qt006}
y_{l}(x) = x^{l+1} e^{-x/2} 
\sum_{p=0}^{n'} \frac{n'! (2l + 1)!}
{(n' - p)!(2l + 1 + p)! p!}
\end{eqnarray}

The discrete values of n' and l, the {\em quantum numbers}
derive from the requirement of regularity for the Laguerre
polynomials \cite{MES69}, the quantum numbers for the ground
state of a hydrogen atom are consequently:

\begin{eqnarray}\label{qt007}
n = n' + l + 1 \qquad n, n', l \in N 
\end{eqnarray}

The discrete energy values are a consequence of the same
condition, and the energy levels are therefore:

\begin{eqnarray}\label{qt008}
        n = \frac{e^2}{\hbar} 
        \sqrt{- \frac{m}{2 E_{n}}} 
        \quad \Rightarrow \quad
        E_{n} = - \left( \frac{e^2}{\hbar}  \right)^2
        \frac{m}{2 n^2}
\end{eqnarray} 

The wave functions of the two lowest states n = 1,2 are
described by:

\begin{eqnarray}\label{qt009}
\psi_{100}(\vec r) &=& \frac{2}{a^{3/2} \sqrt{4 \pi}} e^{- r/a} 
\qquad a = \frac{\hbar^2}{m e^2} \nonumber \\
\psi_{200}(\vec r) &=& \frac{1}{\sqrt{8 \pi}}
\left(1 - \frac{r}{2a}\right) e^{- r/2a} \\
\psi_{210}(\vec r) &=& \sqrt{\frac{1}{4 \pi}}
\frac{r}{2 a} e^{- r/2a} \cos \vartheta \nonumber 
\end{eqnarray}

\subsection{Dynamics of the electron}

As recently demonstrated, the Schr\"odinger equation derives from
the wave equation for the wave function $\psi$, if
external potentials are included. It was established, that its 
formulation in the system at rest -- conveniently chosen to be the
system of the hydrogen nucleus -- requires a time dependent form
$\psi(\vec r,t)$. The same conclusion can be drawn from the time
dependent Schr\"odinger equation, since the development of $\psi$
in the standard model involves the transformation
$\psi(\vec r,t) = \psi(\vec r) e^{- i \omega t}$. 
In this case the fully expanded wave function
of the hydrogen electron for the lowest states will be:

\begin{eqnarray}\label{qt011}
\psi_{100} (\vec r,t) &=& C_{100} e^{-r/a} e^{-i\omega_{1}t}
\nonumber\\
\psi_{200} (\vec r,t) &=& C_{200}\left(1 - \frac{r}{2a}\right)
 e^{-r/2a} e^{-i\omega_{2}t}\\
\psi_{210} (\vec r,t) &=& C_{210} \frac{r}{2a} 
e^{-r/2a} \cos \vartheta e^{-i\omega_{2}t}
\nonumber
\end{eqnarray}

To calculate the states of electron motion, deriving from the
standard solution of the Schr\"odinger equation, we may employ
the wave function Eq. \ref{qt011} and deduce the corresponding
frequencies $\omega$ for an arbitrary radius $r$ within the 
hydrogen shell. For the state (100) we get from:

\begin{eqnarray}\label{qt012}
\left(- \frac{\hbar^2}{r^2 2 m} \frac{\partial}{\partial r}
r^2 \frac{\partial}{\partial r} - \frac{e^2}{r}\right)
e^{-r/a} = \hbar \omega(r)e^{-r/a}
\end{eqnarray}

the frequency $\omega(r)$:

\begin{eqnarray}\label{qt013}
\hbar \omega(r) = \frac{1}{r}
\left(\frac{2 \hbar^2}{2 m a} - e^2\right) - 
\frac{\hbar^2}{2 m a^2}
\end{eqnarray}

As could be exspected from the formulation of Schr\"odinger's
equation, the frequencies depend on the distance from the nucleus.
Setting $\omega(r)$ to zero and calculating the corresponding radius,
the result will be:

\begin{equation}\label{qt014}
r(\omega = 0) = (2 - 2) a = 0
\end{equation}

The consequence of the conventional treatment will be
{\em negative} frequencies and {\em virtual} velocities
of the electron wave. 
Employing a mechanical analogy by asssuming that the electrostatic
potential only describes modifications of the electron wave, 
while kinetic properties are described by:

\begin{equation}\label{qt017}
\hbar \omega(r) = \frac{1}{r} \left(
\frac{\hbar^2}{m a}\right) - 
\frac{\hbar^2}{2 m a^2}
\end{equation}

does not significantly change the result: in this case the
radius, where frequencies become negative and velocities
consequently virtual begins at $2a$. The result points to a
difficulty deterministic theories are generally confronted with:
since the energy of individual waves is generally positive, the
reproduction of negative eigenvalues for any specific wave is
not possible. This feature usually requires to introduce
negative energies by mechanical analogies. But in this case a double
perspective -- from the viewpoint of intrinsic field and wave 
properties as well as external mechanical properties -- is inherent
to every further development of the model, and this double
perspective commonly results in major inconsistencies. The same
applies to the new framework of material wave theory. If the
properties of electrons in the hydrogen system had to be treated on
the basis of electrostatic interactions, then a logically consistent
model could not be achieved. That electrostatic interactions have been
referred to photon exchange and consequently intrinsic field
properties of single particles, provides the decisive difference to
conventional models.  

The calculation of frequencies and physical properties of hydrogen 
electrons can be repeated for the states (200) and
(210) with similar results. It can therefore be concluded,
that the standard ground state solutions of quantum theory 
for the hydrogen atom do not allow for an interpretation of the
wave function as a {\em physical} wave. In the context of the
framework developed, the result can be interpreted in two ways:
(i) The standard solutions of the Schr\"odinger equation
for the hydrogen atom provides the correct eigenvalues, 
but the related wave functions are not valid solutions if seen
from a physical point of view. (ii) The frequency of the electron
within the hydrogen atom is not a valid physical variable. 

The first conclusion requires, that the model of hydrogen atoms
be reformulated in such a way, that the physical properties of
electron waves are accounted for. The second conclusion is only
legitimate, if the Copenhagen interpretation \cite{COP26}
of quantum theory is accepted: but as the framework of material 
waves could already establish, the Copenhagen interpretation has 
to be rejected, because the wave function of a single particle
also possesses physical significance 
\cite{HOF98A}.

\subsection{Theoretical limitations}

The standard model of hydrogen atoms has been remarkably sucessful
in formalizing the experimental results of a wide range of
measurements. The concept of electron orbits, structured by radial
and rotational states of the hydrogen electron in combination with
the intrinsic variable of electron spin allows to account
for all effects related to the lines of spectral emission as
well as their fine--structure. 

The wealth of theoretical and experimental verifications
of the standard theory seems to yield any modification either
fruitless, since it would be contradicting experimental evidence,
or outright senseless, as the theoretical framework is complete
anyway. In view of the results of the preceding section, that
the eigenstates of electrons due to the standard theory are not
valid solutions for {\em physical} waves, the theoretical
development of a framework of physical waves must either stop
short of atomic properties, which limits the range of application
to electron--photon interactions and the description of single
particles, or be extended to account also for atomic properties. 

It must be mentioned that a theory of physical waves
of the hydrogen electron is severly limited compared to the 
standard theory, because every assumption must be backed by
physical plausibility. While quantum theory can easily postulate
the existence of additional parameters, a theory of physical 
waves {\em must} at every step be in accordance with (i) the 
principle of causality, (ii) the basic qualities of particles 
and waves defined by the extended framework of electrodynamics, 
and (iii) the modes of interactions defined by photon emission and
absorption processes. The result of a theoretical development
in this case can be nothing short of a deterministic theory of
hydrogen atoms. 

Apart from a limitation, these constraints of development are
also a theoretical challenge, and it cannot be guaranteed, that
the first attempt in this field will provide the ultimate solution
to the problem. The current paper therefore proposes what is to be
a {\em dynamic theory of hydrogen atoms}, including the basic 
experimental results of spectroscopy as well as a limited 
amount of nuclear physics.

\subsection{Experimental evidence}

Due to its importance for the development of atomic models hydrogen
is by far the best researched atomic system. Experimental data 
relevant for a first dynamic model of hydrogen include the
following:
(i) Thermodynamic data including density at varying 
thermodynamic conditions \cite{HCP76}. (ii)
Spectroscopic data determining the electromagnetic emssion 
frequency and the structure of emission lines, and x--ray
spectroscopy for the measurement of ionization energies
\cite{BOHR13,SPE33,MAC49,KEL87}.
(iii) And finally, nuclear scattering experiments to determine the
dimensions of the nuclear radius \cite{MUS88,SIM80}.

%---------------------------------------------------------------------------%
%      SECTION 3:  W A V E   S T R U C T U R E                              %
%---------------------------------------------------------------------------%
\section{Structure of the hydrogen shell}

As the standard solutions provided in quantum theory do not
account for physical properties of electron waves, the theoretical
model of hydrogen atoms has to be reformulated. Contrary to the
standard model, the electron in material wave theory does not have 
to be point--like, it can be distributed over a finite region of
space. This result suggests to give up the distinction between
the {\em electron} and the region of its 
{\em probability}--distribution, the shell of the hydrogen atom.
Giving up the distinction means, that the electron itself {\em is}
the hydrogen shell, while the internal structure of the shell is
determined by internal properties of the {\em moving} electron.
From a theoretical point of view, the internal structure of
electrons in motion is described by the wave equation for the
wave function $\psi(\vec r,t)$:

\begin{equation}\label{ws000}
        \triangle \psi(\vec r,t) - \frac{1}{u_{el}^2}
        \frac{\partial^2}{\partial t^2} \psi(\vec r,t) = 0
\end{equation}

For the electron wave within the
basically spherical region of the hydrogen shell, the geometrical
features derive from a definite atomic radius and assumed spherical
symmetry. This consideration excludes states of motion based on
orbital rotations, because (i) they would result in unsteady states
along the axis of rotation, and (ii) rotation of electron waves is 
not limited to discrete frequencies, as can be derived from the 
modified de Broglie relation for electron waves \cite{HOF98A}:

\begin{eqnarray}
        \psi (\vec r,t) = \psi_{0} \sin (\vec k \vec r - \omega t)
        \quad
        \vec k \vec r = \frac{m}{\hbar} (\vec \omega \times \vec r)
        \cdot \vec r = 0 \nonumber \\
        \psi = \psi (t) = \psi_{0} \sin (\omega t)
\end{eqnarray}

If motion cannot be lateral motion, then it has to derive from
radial oscillations, and the limiting spheres at $r = R_{N}$, the
radius of the proton, and $r = R_{0}$, the radius of the hydrogen
shell must then be functional nodes. 

\subsection{Resonant states of motion}

These considerations suggest to solve the wave equation for 
radial symmetry of the wave function $\psi$, determining the
allowed states of electron motion by boundary conditions at the
atomic radius $R_{0}$. The solutions for $\psi(\vec r, t)$ then
must be used to determine kinetic properties of the moving electron
and the change of electron states finally has to be referred to the
observed spectral emissions.

Separating the wave function $ \psi = \psi (\vec r, t) $ of electron
waves into a function $ \chi_{R} (\vec r) $ and $ \chi_{t} (t) $,
where $ \chi_{t} (t) $ shall be periodic with $ \nu_{0} $, the
wave equation \ref{ws000} yields:

\begin{equation}\label{ws001}
        \triangle \chi_{R} (\vec r) + 
        \frac{\omega_{0}^2}{u_{el}^2} \chi_{R} (\vec r) = 0
\end{equation}

The significance of $\nu_{0}$, the {\em resonance frequency}
of the hydrogen system, will be shown later on. It derives,
essentially, from photon interactions of the nucleus and the
electron shell.

For a function $ \chi_{R} (\vec r) = \chi_{R} (r) \cdot Y (\vartheta, \phi) $
with $Y(\vartheta, \phi) = 1$, solutions complying with the boundary
conditions at $R_{0}$ are, in the simplest case, spherical Bessel functions
described by:

\begin{eqnarray}
        \chi_{R} (r) = \frac{\sin ( k_{n} r )}{r} \label{ws002} \\
        k_{n} = \frac{\omega_{0}}{u_{n}} \qquad u_{n} = 
        \frac{\nu_{0} R_{0}}{n} \nonumber
\end{eqnarray}

Density $ \rho (r) $ is proportional to $ \chi_{R}^2 (r) $, its
amplitude can be inferred from electron mass $ m_{e} $ within 
the shell:

\begin{equation}\label{ws003}
        \rho (r) = \rho_{el}^{0} \, \frac{\sin^2 (k_{n} r)}{r^2}
        \qquad \int\limits_{V_{s}} d^3 r'\,\rho (r') = m_{e}
\end{equation}

Radial velocities of the electron cannot be generally constant, if 
the hydrogen atom is stable. They must therefore contain time
dependent oscillations, determined by the resonance frequency of
the system. 
Generalizing radial velocity $ u_{n} $ within the shell to a 
function periodic with $ \nu_{0} $, we may write:

\begin{equation}\label{ws004}
        u_{n} (t) := u_{n} \,\cos (\omega_{0} t)
\end{equation}

And the radial momentum $p_{r}(r,t)$ for an arbitrary radius
$r$ and an arbitrary moment $t$ is then given by:

\begin{equation}\label{ws005}
        p_{n} (r, t) : = \rho (r) u_{n} (t) =
        \rho_{el}^{0} \,u_{n} \,\frac{\sin^2 (k_{n} r)}{r^2} 
        \, \cos (\omega_{0} t)
\end{equation}

Provided, the constraints are fullfilled, the kinetic state of
internal variables is determined at every given moment.
The model of hydrogen atoms is therefore deterministic, which
is a consequence of internal wave properties, already established
to exist beyond the level of uncertainty defined by Heisenberg's
relations \cite{HEI27}.

\subsection{Electron wave properties}

The kinetic properties of the radial electron wave will now be
developed for the moment $t = 0$, which will account for the 
features of the electron wave without electron--proton interations. 
Including the effects of interactions will be the subject of the
following section,
since the interplay between the locally periodic structure 
$\sin^2(k_{n} r)$ and the oscillation with constant frequency 
$\cos(\omega_{0} t)$ has to be analyzed separately. 
It will be seen, that the local distribution accounts for the 
energy levels of the electrons, while constant oscillation results 
from electron--proton interactions,

The radius of the radial wave nodes are at 
$r = R_{0}$, wavelength $\lambda_{n}$ increases with 
increasing $n$: the wavelength of the electron is therefore 
higher than the atomic radius. Using de Broglie's relation 
for the radial velocity $u_{n}$, the wavelength will be:

\begin{eqnarray}\label{ws006}
        \lambda_{n} = \frac{h}{m_{e} u_{n}} 
        = \frac{h}{m_{e} \nu_{0} R_{0}}\,n
\end{eqnarray}
 
Applying the dispersion relation $\lambda_{n} \nu_{n} = u_{n}$
the frequency of the radial electron wave will be:

\begin{eqnarray}\label{ws007}
        \lambda_{n} \nu_{n} =  u_{n} \quad \Rightarrow \quad 
        \nu_{n} = \frac{m_{e}}{h}\,\frac{(\nu_{0} R_{0})^2}{n^2}
\end{eqnarray}

The kinetic energy of electron motion at $t = 0$ will be, with
Eq. \ref{ws005} and \ref{ws003}:

\begin{eqnarray} \label{ws008}
        \phi_{K}(r, t=0) = 
        \frac{1}{2} \rho_{el}^0 u_{n}^2 \sin^2(k_{n} r) \nonumber\\
        W_{K}(t = 0) = \frac{1}{2} u_{n}^2 
        \underbrace{\int_{V_{S}} d^3 r\, 
        \rho_{el}^0 \frac{\sin^2(k_{n} r)}{r^2}}_{=\, m_{e}} = 
        \frac{1}{2} m_{e} u_{n}^2
\end{eqnarray}

The kinetic component covers only half of the total energy, 
as could be established by the framework of internal particle 
structures. The additional component, deriving from the
electromagnetic properties of particles \cite{HOF98A}, and
total energy of the electron wave will therefore be:

\begin{eqnarray}\label{ws009}
        W_{E}(t = 0) &=& W_{K}(t = 0) = 
        \frac{1}{2} m_{e} u_{n}^2 \nonumber\\
        W_{T}(t = 0) &=& W_{E} + W_{K} = m_{e} u_{n}^2 \\
        W_{T}(t = 0) &=& m_{e} \frac{(\nu_{0} R_{0})^2}{n^2} 
        = h \nu_{n} \nonumber 
\end{eqnarray}

The level of electron energy in the hydrogen system is proportional
to $n^{-2}$, which is equivalent to the eigenvalues found by the
standard model derived from Schr\"odinger's equation 
(see Eq. \ref{qt008}), although, so far, all energy values are
positive quantities. 

\subsection{Electron--proton interactions}

The model developed so far could neither account for binding 
energy of the electron in the hydrogen system, nor is it a
physically valid solution of the problem. The first is a consequence
of positive energy values, the latter derives from the fact, that
the solution proposed in Eq. \ref{ws005} does not comply with
the wave equation. 

\begin{eqnarray}\label{ws010}
        \left(\triangle - \frac{1}{u_{n}^2} 
        \frac{\partial^2}{\partial t^2}  \right) p_{n}(r,t) 
        =: \Omega(r,t) \ne 0
\end{eqnarray}

Physically, the result signifies, that contrary to motion of a free
electron -- described by the wave equation -- radial electron waves
in the hydrogen system are not self--sustained. The electron wave 
can only prevail due to some kind of interaction accounting for the
variation of intrinsic potentials $\phi_{E}(r,t)$ and $\phi_{K}(r,t)$.
Since the hydrogen system consists only of a nucleus -- the proton --
and the hydrogen shell -- the electron --, interactions must be
electron--proton interactions.

In the standard model the interactions are electrostatic interactions,
and the potential of interaction is given by Coulomb attraction.
By analyzing electrostatic interactions from the viewpoint of internal
particle properties, it was found that electrostatic interactions
can be referred to an exchange of photons, and that photon emission
and absorption processes signify a change of total intrinsic potentials
\cite{HOF98A}. It could equally be derived, that for two interacting
particles the time derivative of emitted and absorbed photon potentials
equals zero. With the total potential of electron motion at an arbitrary
radius $r$ for an arbitrary moment $t$:

\begin{eqnarray}\label{ws011}
\phi_{T}(r,t) = \rho_{el}^0 \frac{u_{n}^2(t)}{r^2} =
\rho_{el}^0 \frac{u_{n}^2}{r^2} \cos^2(\omega_{0} t)
\end{eqnarray}

the time derivative of $\phi_{T}$ yields the changes of intrinsic
potentials due to non--uniform motion. These changes must lead
to photon emission or absorption. If, consequently, the electron 
potential increases, the reason must be an absorption of photons 
emitted by the hydrogen nucleus. The same line of reasoning applies
to a decrease of electron potential: in this case the electron will
be the source of photon emission.
The photon potential due to nuclear interactions can then
be determined. At $r,t$ the photon potential $\phi_{ph}(r,t)$ of
nuclear emission is described by:

\begin{eqnarray}\label{ws012}
        \frac{\partial \phi_{T}(r,t)}{\partial t} =
        - \rho_{el}^0 \frac{u_{n}^2}{r^2} 2 \omega_{0} 
        \sin(\omega_{0} t) \cos(\omega_{0} t) \nonumber \\
        \frac{\partial}{\partial t}
        \left(\phi_{T}(r,t) + \phi_{ph}(r,t)\right) = 0 \\
        \phi_{ph}(r,t) = \rho_{el}^0 \frac{u_{n}^2}{r^2}
        \sin^2(\omega_{0} t) \nonumber
\end{eqnarray}

Due to the spherical setup and accounting for constant energy
flow through sherical surfaces, the total potential of the 
exchanged photon will be, at the nuclear radius $R_{N}$ and 
at the moment $t_{ret} = t \pm \frac{r - R_{N}}{c}$:

\begin{eqnarray}\label{ws013}
        \phi_{ph}(R_{N},t_{ret}) =
        \rho_{el}^0 \frac{u_{n}^2}{R_{N}^2}
        \sin^2(\omega_{0}t_{ret})
\end{eqnarray}

The emitted photon must originate from non--uniform motion of the
hydrogen nucleus, using the same procedure, motion at the nuclear
surface can then be described by:

\begin{eqnarray}\label{ws014}
        \frac{\partial}{\partial t}
        \left(\phi_{nuc}(R_{N},t_{ret}) + 
        \phi_{ph}(R_{N},t_{ret})\right) = 0 \nonumber \\
        \phi_{nuc}(R_{N},t_{ret}) = \rho_{el}^0 
        \frac{u_{n}^2}{R_{N}^2}
        \cos^2(\omega_{0} t_{ret})         
\end{eqnarray}

The hydrogen system, in this case, must be seen as a system of coupled
oscillations, motion of the hydrogen shell is radial and 
non--uniform, while interactions are essentially dynamic processes
accomplished by photon exchange between the hydrogen nucleus and
the electron shell. 

\subsection{Resonance frequency $\omega_{0}$}

The properties of nuclear motion have not yet been determined.
It can be, initially, concluded, that nuclear potentials at $R_{N}$
must be related to non--uniform motion of the nuclear surface. This
point has, since the first version of the model, been clarified. It
was found that oscillation of charge density of the proton is a
sufficient condition for the occurence of oscillating fields. These
fields are transmitted via photons. The only difference, in the
new perspective, is that the whole proton is subjset to density
oscillations, not only its surface. See the publication on
dynamic charge \cite{hof0001012}.
The potential of motion must basically comply with the form:

\begin{eqnarray}\label{ws015}
        \phi_{nuc}(R_{N},t) := \rho_{nuc}^0(R_{N}) u_{nuc}^2(t)
\end{eqnarray}

the result suggests to refer the potential to oscillations of the
nucleus with a frequency of oscillation equal to $\omega_{0}$.
In this case $\omega_{0}$, the resonance frequency of the system,
derives from material properties of the nucleus, while individual
states of radial motion define the state of excitation of the
hydrogen system. From Eq. \ref{ws014} and \ref{ws015} we get:

\begin{eqnarray}\label{ws016}
        u_{nuc}^2 (t) &=& u_{n}^2 \frac{\rho_{el}^0}
        {\rho_{nuc}^0 R_{N}^2} \cos^2(\omega_{0} t) \nonumber \\
        u_{nuc} (t) &=& \pm u_{n} \sqrt{\frac{\rho_{el}(R_{N})}
        {\rho_{nuc}(R_{N})}} \cos(\omega_{0} t)  \\
        u_{nuc} (t) &=& \pm \, \alpha u_{n} \cos(\omega_{0} t)
        \qquad \alpha =  \sqrt{\frac{\rho_{el}(R_{N})}
        {\rho_{nuc}(R_{N})}} \nonumber
\end{eqnarray}

Oscillation of the nuclear surface complies with the harmonic 
equation, as can easily be established:

\begin{eqnarray}\label{ws017}
        r_{N}(t) &=& r_{N}^0 \pm \frac{\alpha u_{n}}
        {\omega_{0}} \sin(\omega_{0} t) \nonumber \\
        \ddot{r}_{N}(t) \rho_{nuc}(R_{N}) 
        &=& - \kappa \cdot (r_{N}(t) - r_{N}^0) \\
        \kappa &=& \rho_{nuc}(R_{N}) \omega_{0}^2 \nonumber
\end{eqnarray}

\subsection{Energy values and atomic radius}

The hydrogen system can therefore be interpreted as a {\em dynamic}
system of radial electron waves and coupled electron--proton 
oscillations, where coupling is effected by photon emission and
absorption processes. The energy values describing a specific state
of excitation $n$ are therefore: (i) the energy of radial motion
of the electron, generally a positive value, and (ii) the energy
contained in electron--proton interactions responsible for the
energy level of coupling. To estimate the binding energy of an 
electron in a state $n$ of excitation, it has to be considered, that
the undisturbed motion of the electron would signify a total
energy of motion equal to $m_{e} u_{n}^2$. Due to the interaction
process, the average energy in one full period is reduced by:

\begin{eqnarray}\label{ws018}
W_{el}^0 &=& m_{e} u_{n}^2 \nonumber \\ 
\langle W_{el} \rangle &=& W_{el}^0 \,
\frac{1}{\tau} \int_{0}^{\tau} \cos^2(\omega_{0} t) dt = 
\frac{1}{2} m_{e} u_{n}^2 \\
W_{int} &=& - \left(W_{el}^0 - \frac{1}{2} m_{e} u_{n}^2\right) =
- \frac{1}{2} m_{e} u_{n}^2  \nonumber
\end{eqnarray}

Contrary to the standard model, the interaction energy of the 
electron is not independent from its state of excitation. For 
low excitation values, corresponding to low energy levels of
the hydrogen system and $n \rightarrow \infty$, the binding
energy of the electron aproaches zero. The physical significance
of this result will be analyzed later on, currently the important
consequence will be, that the physical concept of {\em ionization}
has to be defined differently for the dynamic model of atoms than
is usual in quantum theory. For a specific level of excitation
$n$ the total energy of the electron and the energy of electron
proton interactions are with Eq. \ref{ws002}:

\begin{eqnarray}\label{ws019}
W_{el}(n) &=& m_{e} (\nu_{0} R_{0})^2 \, \frac{1}{2 n^2} \nonumber \\
W_{int}(n) &=& - m_{e} (\nu_{0} R_{0})^2 \, \frac{1}{2 \,n^2}
\end{eqnarray}

For the calculation of the parameters $\nu_{0}$ and $R_{0}$ the 
deduced results have to be related to experimental data. We
define now the {\em ionization} energy of the hydrogen system 
as the binding energy in its highest state of excitation $n = 1$,
its value will be about $E_{0} = - 13.598$ eV.  In this case we get:

\begin{eqnarray}\label{ws020}
        \nu_{0} R_{0} = \sqrt{\frac{- 2 E_{0}}{m_{e}}} = 
        2.187 \times 10^6 [m s^{-1}]
\end{eqnarray} 

The energy value is the energy difference between the
hydrogen ground state ($n \rightarrow \infty$) and the highest
level of excitation and also equal to the energy contained in
electron--proton interactions. Ionization energy therefore can
either be seen as the quantity of energy necessary to break the 
bonds of a hydrogen electron or as the quantity of energy necessary 
to excite the highest state of oscillation.

And the radius of the hydrogen atom can be calculated by 
employing the relation for the wavelength of radial electron
waves Eq. \ref{ws006}, which shall be, for $n = 1$, 
equal to the atomic radius:

\begin{eqnarray}\label{ws021}
        R_{0} = \frac{h}{\sqrt{2 \,E_{0} m_{e}}} 
        = 3.33 \times 10^{-10} [m] \nonumber \\
        \nu_{0} = 6.57 \times 10^{15} [Hz]
\end{eqnarray} 

Comparing with the density of hydrogen in its gas state \cite{HCP76},
the total volume of 5.37 $\times 10^{22}$ H atoms (we assume atomic
hydrogen, the number of atoms equals the mass of 0.0899 g) comes to
8.38 $\times 10^{-3}$ l, or about 1 \% of the total volume of
hydrogen gas in standard conditions.  In its liquid and solid
state the calculated volume of the experimentally measured mass
(70.6 -- 70.8 g) at low temperatures will be 6.59 l, or about six 
to seven times the total volume of hydrogen at the conditions of 
measurements. Accounting for molecular densities by a reduction of 
about 2 still leads to a volume higher than 3 l. The result
indicates that liquid and solid states do not leave the structure
of hydrogen atoms or molecules unaffected. 

\subsection{Internal fields}

The radial and intrinsic momentum of the electron wave and the
complementary electromagnetic potential in the hydrogen shell are:

\begin{eqnarray}\label{ws022}
        \vec p(r,t) &=& \rho_{el}^0 u_{n} 
        \frac{\sin^2(k_{n}r)}{r^2} \cos(\omega_{0}t)\, \vec e^r \nonumber \\
        \phi_{int}(r,t) &=& \rho_{el}^0 u_{n} 
        \frac{\cos^2(k_{n}r)}{r^2} \cos^2(\omega_{0}t)
\end{eqnarray}

Using the definition of electric fields \cite{HOF98A}:

\begin{equation}\label{ws023}
        \bar \sigma \vec E = - \nabla \phi_{int} + 
        \frac{\partial \vec p}{\partial t}
\end{equation}

the change of internal properties can be referred to an external
potential $ - \nabla \phi_{ext} = \bar \sigma \vec E$, which
must be described by:

\begin{eqnarray}\label{ws024}
        \nabla \phi_{ext}(r,t) = \nabla \phi_{int}(r,t) -
        \frac{\partial \vec p}{\partial t}
\end{eqnarray}

Since the setup is spherically symmetric, an integration over
a sphere with radius $r$ yields, with the help of Gauss' theorem:

\begin{eqnarray}\label{ws025}
        \phi_{ext}(r,t) = \phi_{int}(r,t) + 
        \rho_{el}^0 u_{n} \omega_{0} \frac{\sin(\omega_{0}t)}{2 r}
\end{eqnarray}

The amplitude of density can be calculated from Eq. \ref{ws003},
together with the expression for velocity $u_{n}$ from Eq.
\ref{ws002}, the potential will be:

\begin{equation}\label{ws026}
        \phi_{ext}(r,t) = \phi_{int}(r,t) + 
        \frac{m_{e} \nu_{0}^2}{2 n}
        \frac{\sin(\omega_{0}t)}{r}
\end{equation}

The first term is the consequence of electron motion, it is the
{\em self interaction term} of the field within the hydrogen
shell. The second term must be referred to exterior origins,
to the hydrogen nucleus. 

The result can be used to determine the nuclear potential of 
the proton. Since the external potential must result from 
proton energy, and since the interaction by photons already
established harmonic oscillation of the nuclear surface, the
periodic features of the potential suggests a relation with
the harmonic qualities of the nuclear radius. But if the
potential is due to changes of the nuclear radius, the 
nuclear potential can be calculated.

\subsection{Nuclear potential}

The nuclear potential due to a radius $r_{N}(t)$ may
be written:

\begin{equation}\label{ws027}
\phi(r_{N}(t)) := \frac{\phi_{N}^0}{r_{N}(t)}
\end{equation}

And in first approximation the change of nuclear radius will
consequently result in periodic fields described by:

\begin{eqnarray}\label{ws028}
\phi(r_{N}(t)) &=& \frac{\phi_{N}^0}
{r_{N}^0 \pm \frac{\alpha u_{n}}{\omega_{0}} \sin(\omega_{0}t)}
\nonumber \\
\triangle \phi(r_{N}(t)) &\approx& 
\mp \frac{\phi_{N}^0}{(r_{N}^0)^2} 
\frac{\alpha u_{n}}{\omega_{0}} \sin(\omega_{0}t)
\end{eqnarray}

Comparing with the result for the external field in the hydrogen
shell and accounting for radial symmetry of the problem will yield,
for the intensity $\phi_{N}^0$ of the nuclear field, the
expression (nuclear density assumed constant):

\begin{eqnarray}\label{ws029}
\frac{(\phi_{N}^0)^2}{R_{N}^3} = \mbox{const} =
\nu_{0}^2 \frac{3 \pi^2 m_{e} m_{p}}{R_{0}} \nonumber \\
\frac{(\phi_{N}^0)^2}{R_{N}^3} =
5.78 \times 10^{-15} [kg^2/m s^2]
\end{eqnarray}

The analysis of proton potentials can be carried one step 
further, touching the fundamental problem of proton stability.
As the framework of nuclear forces in the current standard
is completely separated from the interactions in atomic or
macro physics, it cannot easily be compared to the nuclear
potentials derived. The dynamic theory of hydrogen atoms,
furthermore, is limited to only two material parameters,
density of mass and density of charge. These limitations
would either require an extension of particle properties in
the hydrogen nucleus -- which is, essentially, what the current
standard of QCD (quantum chromo dynamics) is based on --
or a modification of existing parameters to account for the
extreme properties of nuclear mass. Estimating the nuclear potential
due to gravity or electrostatic interactions, we get for
standard coupling and constant nuclear density with:

\begin{equation}\label{ws030}
\nabla \vec G = - \gamma \rho_{N} \quad
\nabla \vec E = + \frac{1}{\epsilon} \sigma
\end{equation}

for the nuclear potential the result ($grav$ refers to gravitational,
$el$ to electrostatic origins):

\begin{eqnarray}\label{ws031}
\phi_{grav} &=& - \gamma \frac{m_{p}\rho_{N} R_{0} \alpha}
{8 \pi^2 r_{N}} \frac{\sin(\omega_{0}t)}{n r} \nonumber \\
\phi_{el} &=& + \frac{e \sigma_{N} R_{0} \alpha}
{32 \pi^3 \epsilon r_{N}} \frac{\sin(\omega_{0}t)}{n r}
\end{eqnarray}

And since this potential is the origin of forces on the electron
wave given by Eq. \ref{ws025}, the required nuclear radius will be
described by:

\begin{eqnarray}\label{ws032}
\epsilon\,r_{N}^{7/2} &=& \frac{6 e^2}{32 \pi^3 m_{e} \nu_{0}^2}
\sqrt{\frac{2 m_{e} R_{0}}{3 m_{p}}} = 1.38 \times 10^{-48} \nonumber \\
\gamma^{-1}\,r_{N}^{7/2} &=& \frac{6 m_{p}^2}{32 \pi^3 m_{e} \nu_{0}^2}
\sqrt{\frac{2 m_{e} R_{0}}{3 m_{p}}} = 4.70 \times 10^{-63}
\end{eqnarray}

From the known coupling constants the nuclear radius can be
calculated. It would be for electrostatic or gravitational coupling:

\begin{eqnarray}\label{ws033}
r_{N}(grav) = 3.69 \times 10^{-21} m \nonumber \\
r_{N}(el) = 3.05 \times 10^{-11} m 
\end{eqnarray}

The nuclear radius, in case of gravity couplings, would be far too
small, in case of electrostatic coupling far too big to be consistent
with nuclear scattering experiments. It can therefore be excluded, that
nuclear stability can be the result of any usual field we deal with in
quantum theory or electrodynamics. The same conclusion has
been drawn in the standard theory, and the consequence was the
development of theories on nuclear forces as well as the extended
framework of QCD. 

\subsection{General solutions}

The solution for radial motion of the electron wave in the hydrogen
atom applied spherical symmetry. In a more general picture, this
form of motion of the electron wave and consequently of the nuclear
surface will only be the simplest form of motion, the monopole
oscillation of the nuclear surface. In a more general picture the
nuclear surface is an oscillating multipole, which must affect
the electron wave in such a way, that the lateral components
of motion, the spherical harmonics $Y_{lm}(\vartheta, \varphi)$
are no longer constant. The constraints for a general solution of
the problem then have to be accounted for in such a way, that the
wave equation for the intrinsic electron momentum $\vec p(\vec r,t)$
is satisfied due to the interactions with nuclear potentials. The
system again has to be a system of coupled oscillations, and the
energy levels can be determined by calculating, in the same manner,
kinetic energy components and components of interaction.    

\subsection{Summary}

In this section we solved the wave equation of the hydrogen electron
for spherical symmetry. The resulting states of radial motion could
be related to the energy levels of excited hydrogen by including the
experimental value of ionization energy. The interaction energy of
nuclear oscillation and electron motion accounted for the periodic
feature of electron oscillation, the binding energy of an electron
was found half of the energy contained in its radial motion.
Internal fields of electron--proton interaction were referred to
nuclear potentials, and by estimating the effects of electromagnetic
and gravitational coupling it could be established, that intrinsic
nuclear interactions cannot be explained by this type of interaction.
We equally showed that the radially symmetric solution is but the
simplest case, and that nuclear multipole oscillation provides
additional modes of electron--proton interactions as well resonant
states of electron motion. 

\section{Photon emission}

\subsection{Balmer's relation}

The energy levels of electron motion within the hydrogen shell
account for the discrete frequency of spectral emission, if the
change of energy from a level $n$ to a level $m$ is emitted
as photon energy, since with \ref{ws018} the energy difference
must be:

\begin{equation}\label{em001}
\triangle W_{nm} = \frac{1}{2} m_{e} (u_{n}^2 - u_{m}^2) =
\frac{m_{e}}{2} (\nu_{0}R_{0})^2 \left(
\frac{1}{n^2} - \frac{1}{m^2}\right)
\end{equation}

And using Planck's relation for the frequency of the
emitted photon, the result will be Balmer's relation:

\begin{eqnarray}\label{em002}
\nu_{nm} &=& \frac{m_{e}(\nu_{0}R_{0})^2}{2 h}
\left(\frac{1}{n^2} - \frac{1}{m^2}\right)
= \nu_{em}^0 \left(\frac{1}{n^2} - \frac{1}{m^2}\right)
\nonumber \\
\nu_{em}^0 &=& 3.288 \times 10^{15} Hz
\end{eqnarray}

The result, although defining the allowed emission frequency of
photons, does not provide any insight into the process of 
emission. Within the framework of a dynamic model of atoms it is
therefore only half of the whole story. While induced emission,
the transition due to external origins, can be seen as a process
determined by interactions, the same does not apply to spontaneaus
emissions, which must be referred to internal origins. If 
spontaneous emissions occur, then the theoretical framework must
provide a model which meets (i) the properties derived for the
intrinsic variables, and (ii) the requirements of thermodynamics
if a statistical ensemble of atoms is considered. 

That spontaneous emissions occur is suggested by the principles
of thermodynamics, which require that entropy in natural processes
increases, which means, that atoms with a high level of excitation
must in a closed system emit their energy until thermodynamic
equilibrium, corresponding to the maximum of entropy, is achieved.
We account for these requirements by a model, which limits
statistical considerations to inter--atomic processes, while
the process of emission itself shall be deterministic.

\subsection{Causality and probability}

The transition of electron motion from a level $n$ to a level
$m$ in the standard theory involves selection rules and probability
considerations. The procedure reflects the interpretation of 
electron waves as probability waves and is a conesequence of the
fundamental assumptions in quantum theory, as already demonstrated
\cite{HOF98A}.

In the framework of material wave theory, events and processes on
the level of electron waves are described, although only partly,
by Maxwell's equations, the events are therefore determined. Which
means, that probability cannot enter at this level of theoretical
formulations, but must be accounted for on the level of thermodynamic
environments. The emission of photons as well as the transition from
one state $n$ to a state $m$ is therefore related to probability only
insofar it refers to the mechanical and electromagnetic interactions
of one atom with other atoms. Excitation and emission then are
processes governed by the general rules of Maxwell--Boltzmann
distributions, where the probability of a state $E_{n}$ of atomic
excitation can be described by:

\begin{eqnarray}\label{em011}
W(n) \propto e^{- E_{n}/k_{B}T}
\end{eqnarray}

\begin{figure}
\epsfxsize=0.9\hsize
\epsfbox{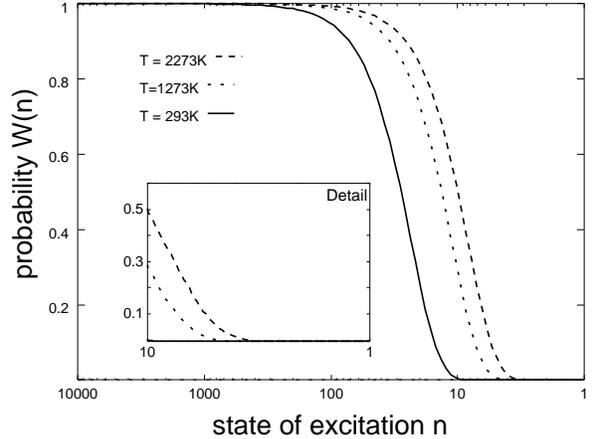}
\caption{Probability of states of excitation $n$ at standard
conditions 293K, at 1273K and at 2273K. The probability of
excited states with $n < 10$ is insignificant compared to
states with $n > 100$}\label{fig004}
\end{figure}

The probability of higher levels of excitation is generally very 
low, most hydrogen atoms will be in a state $n > 100$ of excitation.
The result suggests, that spectroscopic data based on the frequency
range of visible or near visible emisssion, should concern only
an insignificant minority of physical processes in a given 
environment. The probability of individual states $n$ has been
plotted in Fig. \ref{fig004}, it can be seen, that the transition
interval from probability 1 to probabilities close to 0 in standard
conditions encompasses the range of states from $n = 10$ to 
$n = 100$.

That interactions of the atom with its environment are probabilistic
does not imply, though, that the transitions internally are also
subject to the same consideration. It can equally mean, that the
internal processes are strictly determined, once the initial energy
value is acquired by inter--atomic processes. We develop the
consequence of this concept for the process of photon emission by
modifying the relations of electron and nuclear excitation to account
for decay processes. Since the hydrogen atom is a coupled system of
nuclear and electron oscillations, decay of either electron motion
or nuclear oscillation will have the same effect: the level of
excitation is reduced.

\subsection{Deterministic model}

Interactions between nuclear oscillations and radial electron waves
were the origin of internal stability of excited atoms. The
problem of spectral emission of non-uniform motion
can be treated in a deterministic way by an additional assumption:
nuclear oscillations shall not be constant, but subject to decay.
Then the transition from resonant state n $ \rightarrow m $ can be
based on the formulation for nuclear velocity $ u_{nuc}^2 (t) $:

\begin{equation}\label{em012}
\rho_{nuc}^0 \, u_{nuc}^2 (t) = \rho_{el}^0 \, u_{n}^2 \, e^{- \alpha t}
\qquad \alpha = \frac{2}{\tau_{\epsilon}} \ln \frac{m}{n}
\end{equation}

where $ \tau_{\epsilon} $ is the emission interval. 
If nuclear motion is altered, the stability of radial motion is no
longer sustained, and the time derivative of total potential at any
arbitrary point of the shell is not zero:

\begin{equation}\label{em013}
\frac{\partial \phi_{T}^{nuc}}{\partial t} +
\frac{\partial \phi_{T}^{el}}{\partial t} = 
\frac{\partial \phi_{T}}{\partial t} \ne 0
\end{equation}

We assume now, that the time derivative of total potential is
emitted in the form of photons. Photon energy density can then be 
calculated, and integrating over the atomic shell will allow to 
determine the frequency of spectral emission.
Setting total potentials of nuclear and electron motion equal to:

\begin{eqnarray}\label{em014}
\phi_{T}^{nuc} (r, t) &=& \frac{\rho_{el}^{0} u_{n}^2}{r^2}
\sin^2 (\omega_{0} t) \,e^{- \alpha t} \nonumber \\
\phi_{T}^{el} (r, t) &=& \frac{\rho_{el}^{0} u_{el}^2 (t)}{r^2}
\cos^2 (\omega_{0} t)
\end{eqnarray}

The time derivatives of total potential and consequently
the total potential of emission at r  yields a non--linear
differential equation, which
can be transformed into a linear one
by  substituting $ z (t) = u_{el}^2 (t) $:

\begin{equation}\label{em015}
z' (t) + z (t) \, \frac{g (t)}{f (t)} = - \frac{h (t)}{f (t)}
\end{equation}

\begin{eqnarray*}
f (t) & = & \cos^2 (\omega_{0} t) \qquad g (t) = 
2 \omega_{0} \cos (\omega_{0} t) \sin (\omega_{0} t) \\
h (t) & = & u_{n}^2 e^{- \alpha t} \left(
2 \omega_{0} \cos (\omega_{0} t) \sin (\omega_{0} t) - \alpha
\sin^2 (\omega_{0} t) \right) + \\
&+& \frac{r^2}{\rho_{el}^{0}}
\frac{\partial \phi_{T}^{ph}}{\partial t}
\end{eqnarray*}

In first approximation, assuming the time derivative of photon 
potential constant, the general solution will be:

\begin{eqnarray}\label{em016}
z (t) &=& \frac{1}{M (t)} \left\{
- \int \frac{h (t)}{f (t)} M (t) dt + C \right\} \nonumber \\
M (t) &=& \exp \left( \int \frac{g (t)}{f (t)} dt \right)
\end{eqnarray}
which yields the equation for electron velocity:

\begin{eqnarray}\label{em017}
u (t) & = & \frac{1}{\cos (\omega_{0} t)} 
\sqrt{C - u_{n}^2 e^{- \alpha t} \sin^2 (\omega_{0} t) - 
\frac{r^2}{\rho_{el}^{0}}
\, \frac{\partial \phi_{T}^{ph}}{\partial t} \, t} = \nonumber \\
& = &
\frac{1}{\cos (\omega_{0} t)} 
\sqrt{C - \frac{r^2}{\rho_{el}^{0}}
\left(\phi_{T}^{nuc} (r, t) -
\, \frac{\partial \phi_{T}^{ph}}{\partial t} \, t
\right)}
\end{eqnarray}

The constant C can be calculated from the intial value of u (t), while
the value for $ t = \tau_{\epsilon} $ allows to determine photon energy:

\begin{eqnarray*}
u^2 (0) &=& C = u_{n}^2 \\
u^2 (\tau_{\epsilon}) &=& u_{n}^2 - 
\,\frac{r^2}{\rho_{el}^{0}} \frac{d \phi_{T}^{ph}}{d t} 
\tau_{\epsilon} = u_{m}^2
\end{eqnarray*}

\begin{equation}\label{em018}
\frac{\partial \phi_{T}^{ph}(r, t)}{\partial t} \, \tau_{\epsilon} = 
\phi_{T}^{ph} (r) = \frac{\rho_{el}^{0}}{r^2}
\left(u_{n}^2 - u_{m}^2 \right) \,
\end{equation}

And substituting the variable $ x = t/\tau_{\epsilon} \quad x \in [0,1] $, 
the potential of electron motion during emission which, for the sake of
simplicity, is taken to prevail for $ \tau_{0}/2 $ seconds:

\begin{eqnarray}\label{em019}
\phi_{T}^{el} (x, r)  &=& 
\frac{\rho_{el}^{0}}{r^2}u^2 (x) \cos^2 (\pi x) = \nonumber \\
&=& - \phi_{T}^{nuc} (x, r) +  \frac{\rho_{el}^{0}}{r^2} \left(
u_{n}^2  - \left(u_{n}^2 - u_{m}^2\right) \cdot x\right) 
\end{eqnarray}

A simple integration over the electron mass within the atomic
shell then yields the frequency and energy of photon emission:

\begin{eqnarray}\label{em020}
E_{nm}^{ph} = h \nu_{nm}^{ph} = \frac{m_{e}}{2}
 \left(u_{n}^2 - u_{m}^2 \right)
\nonumber \\
\nu_{nm}^{ph} = \frac{m_{e} (\nu_{0} \, R_{0})^2}{2 h}
\left(\frac{1}{n^2} - \frac{1}{m^2} \right) 
\end{eqnarray}

which is exactly the result already derived and
which is equal to Balmer's law.

\begin{figure}
\epsfxsize=1.0\hsize
\epsfbox{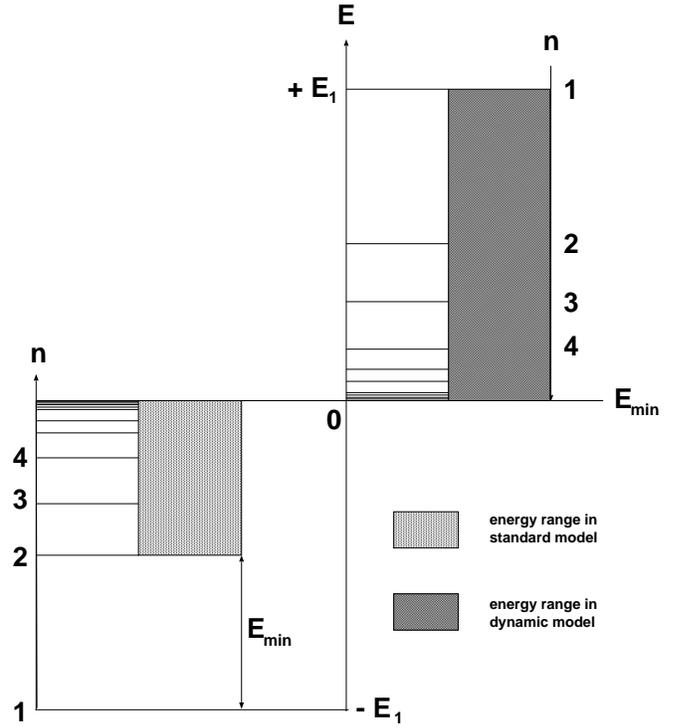}
\caption{Energy terms and interaction energy due to the principal
quantum numbers $n$ in quantum theory (left) and material wave
theory (right). Total energy at the level $n = \infty$ in both
cases equals zero. Total energy in quantum theory is generally
negative, in material wave theory positive. The interaction energy
in quantum theory must be higher than $\triangle E = E_{2} - E_{1}$,
while in material wave theory it can be arbitrarily low
}\label{fig003}  
\end{figure}

The actual dynamical reason leading to exponential decay of nuclear
oscillation is well beyond the scope of this paper. 
But if nuclear oscillations are not stable, then there seems to exist
no reason, why emission shall not continue in the following interval.
If the parameter of nuclear damping $ \alpha $ remains constant, then
the emission interval will in general not be equal to 
$ (2k + 1) \tau_{0}/2 $, but different for every single emission
process. Equally will different decays $ n \rightarrow m_{1} $ and
$n \rightarrow m_{2} $ correspond to different emission intervals:

\begin{equation}\label{em021}
\frac{\tau_{\epsilon_{1}}}{\tau_{\epsilon_{2}}} =
\frac{\ln m_{1} - \ln n}{\ln m_{2} - \ln n}
\end{equation}

Excited atoms will thus lower their level of excitation by continuous
emissions until they are again excited. The process of excitation and
emission - whereby excitation does not seem as easily accessible to
formalisation - is then a basically statistical problem, and while
the emission process is purely deterministic, the interactions with
the environment are not.

\subsection{Thermodynamic environment}

The essential difference between the standard model of hydrogen
atoms and the current one concerns the range of allowed energies.
While hydrogen atoms in quantum theory generally have a negative
level of total energy, the energy in the current model is generally
positive. And while the hydrogen ground state in quantum theory is
described by $n = 1$, the ground state in the current model is
given by $n = \infty$, which describes the level of absolute inertia,
the point where $T = 0 K$ (see Fig. \ref{fig003}). 

Therefore every level of atomic excitation up to the level $n = 1$
allows for electromagnetic emissions, and hydrogen emissions are 
bound to occur well below the visible or infrared spectrum. The 
transition depends, as previously calculated, on the emission 
interval, which should be affected by the enviroment of an emitting
hydrogen atom. The frequency of emission is given by allowed states
of radial motion. Statistical considerations, although without
any relevance to emission frequencies, therefore enter the picture in
the calculation of emission intensities of a specific line of the
hydrogen spectrum. 

%---------------------------------------------------------------------------%
%      SECTION 6:  D I S C U S S I O N                                      %
%---------------------------------------------------------------------------%

\section{Conclusion}

We have analyzed, in this paper, whether the standard solution of
the hydrogen atom, given in quantum theory, is compatible with the
dynamic and intrinsic properties of electrons.
From this analysis it was concluded that the standard model is
essentially invalid, since it does not present a physical (as 
opposed to an only {\em algorithmic}) solution to the problem of hydrogen
spectra. An alternative and physical model of electron waves within
the hydrogen atom was developed, it was shown compatible with the
experimental results of hydrogen spectra. The model reversed the
standard scaling of the electron energies, since the state of
$n = 1$, which is identified as the hydrogen ground state in quantum
theory, is the state of highest excitation in microdyanmics. The
energy of a hydrogen atom depends on the interactions with the
environment and is generally positive. A deterministic and local
picture of the emission process could be deduced, which is based
on the decay of nuclear oscillations. It could also be shown that
the nuclear potential cannot be of electromagnetic or gravitational
origin, which seems to necessitate a separate and different mode
for the interactions in a nuclear environment.

\section{Discussion}

The model of hydrogen, developed in this paper, reveals three interesting
differences to the standard model: 

\begin{itemize}
\item  The radius is not changed if the
atom is excited, contrary to the standard conception of a Rhydberg atom.

\item The liquid and solid state is not merely a combination of two 
identical atoms, but affects the structure and hence the physical 
qualities of single atoms decisively. 

\item There is no minimum excitation
energy: although the next higher state of excitation in a given environment
requires a discrete amount of energy, this amount is not independent of
the thermodynamic environment.
\end{itemize}

These changes affect the concept of hydrogen atoms substantially. Instead
of well-defined and seemingly persistent structures the dynamic atoms not
only change their electron-proton interaction energy with their state of
excitation, but also their very structure in any liquid or solid
aggregation with other atoms. From a rather philosophical point of view
this indicates a departure from the very notion of the atom (in its
original conception an undividable and unchanging entity). The validity
of this new model has to be subjected to experimental tests before it
can become part of physical reality, and the measurements to be
performed are currently developed. Apart from this question the
possibility to formulate a dynamic and hence causal model of hydrogen
atoms - whether this first model will ultimately be suitable or
not - indicates, in a way, a major breakthrough from the purely
statistical description of quantum mechanics.

From the viewpoint of thermodynamics the model seems, on first glance,
better suited to describe the third principle. The energy differences
possess a vanishing low energy limit, whereas the conventional atom in 
quantum theory is characterized by vanishing high energy limits. How 
this feature translates into experimental results has to be worked out. 
But it could, eventually, provide a test to differ between the results 
of microdynamics and quantum mechanics.

%---------------------------------------------------------------------------%
%      SECTION 7:  R E F E R E N C E S                                      %
%---------------------------------------------------------------------------%

%---------------------------------------------------------------------------%
%     E N D  O F  A R T I C L E                                             %
%---------------------------------------------------------------------------%

\end{document}